\documentclass[12pt,preprint]{aastex} 

\usepackage{natbib}
\usepackage{epstopdf}
\usepackage{graphicx}
\usepackage{subfigure}
\usepackage{rotating}
\usepackage{textcomp,graphicx,amsmath,pdflscape,amssymb}

\begin{document}

\def\Msun{M_\odot}

\def\Lsun{L_\odot}

\def\Rsun{R_\odot}

\slugcomment{Submitted to ApJ}

\shorttitle{Ultrafast Novae}

\shortauthors{Shara et al}

\title{A {\it Hubble Space Telescope} Survey for Novae in M87. II. Snuffing out the Maximum Magnitude - Rate of Decline Relation for Novae as a Non-Standard Candle, and a Prediction of the Existence of Ultrafast Novae\altaffilmark{1}}

\author{Michael~M.~Shara\altaffilmark{2,3}, Trisha Doyle\altaffilmark{2,4}, Tod R. Lauer\altaffilmark{5}, David~Zurek\altaffilmark{2}, Edward A. Baltz\altaffilmark{6}, Attay Kovetz\altaffilmark{7}, Juan P. Madrid\altaffilmark{8}, Joanna Miko\l{}ajewska\altaffilmark{9}, J.D. Neill\altaffilmark{10}, Dina Prialnik\altaffilmark{11}, D. L. Welch\altaffilmark{12}, and Ofer Yaron\altaffilmark{13}}

\altaffiltext{1}{Based on observations with the NASA/ESA {\it Hubble Space
Telescope}, obtained at the Space Telescope Science Institute, which is
operated by AURA, Inc., under NASA contract NAS 5-26555.}

\altaffiltext{2}{Department of Astrophysics, American Museum of Natural History, Central Park West and 79th Street, New York, NY 10024-5192, USA}

\altaffiltext{3}{Institute of Astronomy, University of Cambridge, Madingley Road, Cambridge CB3 0HA, United Kingdom}

\altaffiltext{4} {Department of Physics and Space Sciences, Florida Institute of Technology, Melbourne, FL 32901, USA}

\altaffiltext{5} {National Optical Astronomy Observatory, P.O. Box 26732, Tucson, AZ 85726, USA}

\altaffiltext{6} {KIPAC, SLAC, 2575 Sand Hill Road, M/S 29, Menlo Park, CA 94025, USA}

\altaffiltext{7} {School of Physics and Astronomy, Faculty of Exact Sciences, Tel-Aviv University, Tel Aviv, Israel}

\altaffiltext{8} {CSIRO, Astronomy and Space Science, P.O. Box 76, Epping, NSW 1710, Australia}

\altaffiltext{9} {N. Copernicus Astronomical Center, Polish Academy of Sciences, Bartycka 18, PL 00-716 Warsaw,
Poland}

\altaffiltext{10}{California Institute of Technology, 1200 East California Boulevard, MC 278-17, Pasadena CA 91125, USA}

\altaffiltext{11}{Department of Geosciences, Tel Aviv University, Ramat Aviv, Tel Aviv 69978, Israel}

\altaffiltext{12} {Department of Physics \& Astronomy, McMaster University, Hamilton, L8S 4M1, Ontario, Canada}

\altaffiltext{13} {Department of Particle Physics and Astrophysics, Weizmann Institute of Science, 76100 Rehovot, Israel}

\begin{abstract}
The extensive grid of numerical simulations of nova eruptions of \citet{yar05} first predicted that some classical novae might deviate significantly from the Maximum Magnitude - Rate of Decline (MMRD) relation, which purports to characterise novae as standard candles. \citet{kas11} have announced the observational detection of an apparently new class of faint, fast classical novae in the Andromeda galaxy. These objects deviate strongly from the MMRD relationship, exactly as predicted by \citet{yar05}. \citet{sha16} recently reported the first detections of faint, fast novae in M87. These previously overlooked objects are as common in the giant elliptical galaxy M87 as they are in the giant spiral M31; they comprise about 40\% of all classical nova eruptions and greatly increase the observational scatter in the MMRD relation. We use the extensive grid of nova simulations of \citet{yar05} to identify the underlying causes of the existence of faint, fast novae. These are systems which have accreted, and can thus eject, only very low mass envelopes, of order $10^{-7} - 10^{-8} M_{\odot}$, on massive white dwarfs. Such binaries include, but are not limited to, the recurrent novae. These same models predict the existence of ultrafast novae which display decline times $t_2$ as short as five hours. We outline a strategy for their future detection.
\end{abstract}

\keywords{M87, novae, cataclysmic variables}

\section{Introduction and Motivation}
Most astronomers connect Edwin Hubble with the year 1929 because of his momentous
paper ``A Relation between Distance and Radial Velocity among Extra-Galactic Nebulae"
\citep{hub29}, which initiated the study of modern cosmology. In the same month, however,
Hubble also published ``A Spiral Nebula as a Stellar System, Messier 31" \citep{huc29},
in which he announced the resolution of the outer spiral arms of that galaxy into swarms of
faint stars; the discovery of Cepheids and long-period variables; and 63 novae. The study
of extragalactic stellar populations thus began at the same time as cosmology, and classical
novae have played a significant role in populations studies ever since.

Novae are all binaries in which a white dwarf (WD) accretes matter from a hydrogen-rich
brown dwarf, red dwarf or red giant companion; or helium from a white dwarf companion.
When sufficient mass is accumulated that degenerate electron pressure at the base of the
accreted envelope exceeds a critical value, a thermonuclear runaway (TNR) occurs, which
ejects most of the envelope and brightens the WD to its Eddington luminosity or even
brighter \citep{sha81}. Novae near maximum light range in luminosity from M = -6 to -10 \citep{war95}. 
In contrast, the bright end of the planetary nebula (PN) luminosity
function only reaches M(5007) = -4.5 \citep{cia89}; red giant branch
(RG) stars typically reach M = -3 \citep{baa44,san71}; and RR Lyrae stars
achieve $M_V$ = 0.6 \citep{chr66}. Novae can therefore be detected much more easily in a
given galaxy or cluster, and can be observed to significantly greater distances in the field 
than PN, RG or RR Lyrae stars. The transient nature of novae and their 
H$\alpha$ brightness help eliminate contamination due to background emission line objects or unresolved compact
galaxies. 

\citet{zwi36} was the first to announce that novae appeared to behave as standard
candles, with light curves that could be calibrated to yield the distances to galaxies. Zwicky's
first formulations of the MMRD correlation were improved upon by \citet{mcl45} and \citet{arp56}. The physics
of the apparently tight correlation between nova absolute magnitude at maximum light and
their rates of brightness decline was explained by \citet{sha81} and \citet{liv92}. The key
prediction of those investigations is that, all other things being equal, the mass of the white dwarf
in a nova binary is the dominant parameter controlling the behaviours of nova explosions.

The essential physics underlying this prediction is as follows. The degenerate equation of state of matter 
in a WD determines that as the mass of a WD increases, its radius decreases \citep{cha31,cha35}. 
Thus the acceleration of gravity at a WD's surface increases sharply as its mass increases. 
A strongly increasing gravitational potential, with increasing WD
mass, means that much less hydrogen can be accreted onto the WD before a TNR occurs
\citep{sha81}. Lower mass envelopes can be ejected faster than those of higher mass,
so novae occurring on massive WDs will exhaust their thermonuclear-powered envelopes,
and decline in brightness faster, than those on low mass WDs. If WD mass was the {\it only}
free parameter then novae would be luminous, well-understood standard candles displaying
negligible scatter. 

Of course, all other things are {\it not} equal, and novae are decidedly {\it not} a phenomenon
governed by just one free parameter \citep{sta75,sha80}. While it is widely recognized that WD mass is a critical factor
affecting nova explosions, as noted above, it is now understood there there are at least four other important factors that determine
the properties of a nova outburst. These are the accretion rate onto the WD and the
resulting envelope mass \citep{pri82}; the WD luminosity \citep{pri95,yar05};
its chemical composition (He, CO or ONe), and the chemical composition of the accreted
matter (H-rich or He) \citep{fau72, kov85,sta86}. Just the WD mass, accretion rate/envelope mass 
and luminosity can and do produce a rich variety of nova eruptions \citep{pri95,yar05} and scatter about the
so-called MMRD relation. Referring to their nova models, \citet{pri95} stated
that ``Correlations are obtained between the peak luminosity and time of decline...It is shown
that these correlations cannot be tight...The implication is that novae cannot be considered
accurate distance indicators". Do observations bear out this prediction?

Early attempts to measure distances of nearby galaxies, and even to deduce the Hubble constant via novae, had a
reasonably good track record. \citet{cap89} used \citet{coh85}'s calibration of Galactic novae to obtain a distance modulus for M31 
of $24.27 \pm 0.2 $, in good agreement with (m-M) = $24.46 \pm 0.10 $ recently obtained by \citet{deg14}. \citet{cap90} found a distance to the LMC of (m-M) = $18.7 \pm 0.2$, in equally good agreement with the modern value of $18.48 \pm 0.10$ \citep{inn16}. \citet{del95} used the M31 and LMC MMRDs to obtain a Virgo cluster distance of 18.6+-3.3 Mpc, which exceeds by 13\% the modern distance of $16.4 \pm 0.5 $ Mpc \citep{bir10}. Finally, using a sample of just seven novae, a value of the Hubble constant of $70 \pm 13$ km/s/Mpc was obtained by \citet{van92}. \citet{del95} summarised these studies, stating that novae can be used judiciously, 
when geometric and nebular parallaxes are not available, with roughly 30\% errors in distance measurements, 
to non-recurrent Galactic novae. In the modern era of precision cosmology novae are not competitive with much more precise values 
of the Hubble constant obtained via type Ia supernovae (e.g. \citet{rie11}), but 25 years ago the MMRD seemed to be a much better 
distance indicators than the pessimistic assessment of \citet{pri95} then indicated.

Doubts about the MMRD relation were first raised by \citet{fer03}, who noted that:
 
``We examine the maximum magnitude versus rate of decline (MMRD) relation for novae in M49, 
finding only marginal agreement with the Galactic and M31 MMRD relations."  Up to six of the nine novae
detected with the Hubble Space Telescope (HST) in this study appear to be anomalously faint for their fast rates of decline, but conclusive
maxima were only seen for three of the nine novae.

A similar conclusion was reached the following year by \citet{hea04} who, on the basis of 4 well-observed fast novae in the LMC concluded:

``The weighted mean distance modulus to the LMC based on these novae is $18.89 \pm 0.16$. This differs significantly from the distance modulus adopted by
della Valle \& Livio of 18.50...The evidence based on these novae suggests that... some novae in the LMC, including these four, are significantly 
underluminous at maximum light compared with those in M31, by about 0.4 mag".

The strongest recent objection to MMRD came when \citet{kas11} achieved a major breakthrough with their monitoring of M31 for
novae and the resulting observational discovery of ``faint, fast novae". Their nightly cadence
(except when interrupted by weather) and relatively deep magnitude limit overcame the
observational bias against the discovery of such faint, fast transients, inherent in all previous
nova surveys. Rather than being rare outliers, these novae were a significant fraction of all M31 novae detected.
In the past year \citep{sha16} have shown that  these faint, fast novae are as common in M87 as they are in M31.

In section 2 we summarise observations of well-observed novae in the Milky Way, LMC, M31, M33
and M87. We plot the maximum luminosities of novae in these five galaxies versus $t_2$, the
time to decline 2 magnitudes, in section 3, showing that the MMRD should be
discarded as a distance indicator. In section 4 we use the extensive grid of nova models in
\citet{yar05} to explain the observed large observational scatter in the MMRD, and
determine which nova parameters give rise to faint, fast novae. We predict the existence of
ultrafast novae with $t_2$ \textless 1 day in section 5. Our conclusions are summarised in section 6.

\section{Observations}

A compilation of the peak magnitudes and distances to, and hence peak luminosities
of 28 Galactic novae is given by \citet{dow00}. These authors used ground-based and 
HST images of shells, and a mix of their own and literature spectroscopic expansion velocities, 
to determine expansion parallax distances to the largest, uniformly analysed sample of Milky Way novae in the literature. 
We adopt their absolute magnitudes and $t_2$ times to decline from maximum brightness for Milky Way novae,
and add the Galactic symbiotic nova T CrB because of its equally well determined absolute
magnitude (see below). Uncertainties in interstellar reddening are the greatest uncertainty in 
the \citet{dow00} study. This uncertainty adds vertical (magnitude) scatter to the data, but it
cannot selectively hide faint-fast novae.

The then state-of-the-art photographic studies of the LMC were summarised in \citet{cap90}. 
Only 4 novae, at that time, had well-defined (i.e. directly observed, and NOT extrapolated or guessed at) 
times and magnitudes at maximum light. That entire sample, including the large majority of novae with
extrapolated maximum magnitudes and rates of decline, did not detect faint, fast novae. As already noted,
\citet{hea04} expressed doubts about the MMRD on the basis of new observations of fast LMC novae.
The most recent summary of LMC novae is that of \citet{sha13}. Four more novae with well defined times of maximum (within 2 days),
maximum magnitudes and decline times have been observed in the 23 years since \citet{cap90}. These 8 well-observed LMC novae are included 
in our figures described below.

The only long baseline, high cadence, CCD-based survey of the Magellanic Clouds is that of \citet{mro16}. Five  of their 
15 novae with extremely well-defined decline times fall in the faint-fast regime (particularly LMCN 2010-11a and 
 LMCN 2012-03a). Unfortunately their CCD saturates in the magnitude range 11-12, depending on seeing (Mroz 2016, private communication). 
To be conservative we do not include the \citet{mro16} data in our figures below.

The then state-of-the-art photographic studies of M31 novae were summarised in \citet{cap89}.
Unfortunately, the original photometry has not been published, so it is impossible to judge how far 
they have extrapolated the maximum magnitudes, or how well-determined are the rates of decline.
Faint, fast novae are absent from the data. The \citet{sha11} spectrographic survey of M31 novae summarised the previous
decade's photometry of the best studied objects. We include 11 novae with well-defined V-band maxima and $t_2$ in our figures. 
\citet{kas11} used the robotic Palomar 60-inch telescope to sample M31 in single (g) filter images 
in 2008 and 2009, with high cadence, and spectroscopically confirmed several of the transients they discovered as classical novae. 
We adopt \citet{kas11}'s ``best-observed" sample of six faint, fast novae for comparison with our own 
HST observations of M87 novae.

We carried out daily {\it Hubble Space Telescope}/Advanced Camera for Surveys (HST/ACS)
imaging of the giant elliptical galaxy M87 in the F606W (V band) and F814W (I band)
filters taken for HST Cycle-14 program 10543 (PI - E. Baltz) over the 72 day interval 24
December 2005 through 5 March 2006, with a few 5-day gaps at the end of the run. Full
details of the observations, data reductions, detections and characterisations of 32 certain
and 9 likely novae are given in \citet{sha16}. Figures 1 and 2 of that paper include the daily 
images of each nova, and their full light and color curves, respectively. This
survey for extragalactic novae is unprecedented, because HST observations rule out gaps
due to weather, and there are no variations in limiting magnitude due to variable seeing or
lunar phase. Thus 21 novae were detected both before and after maximum light, and their
brightnesses were measured within 12 hours of maximum light. Our daily sampling over a
10 week span was deep enough to be almost impervious to M87's background light, revealing
novae to within 10" of the galaxy's nucleus. In addition, novae were detected over a nearly
6 magnitude range of brightness, so that even the faintest and fastest of novae were easily
detected.

\section{Milky Way, LMC, M33, M31 and M87 MMRD data}

In Figure 1 we plot the MMRD diagram of all the Galactic novae with expansion parallax
distances from the \citet{dow00} study of novae. To these we add T CrB, the
symbiotic nova with a similarly reliable distance and absolute magnitude. For T CrB, $t_2$
is taken from \citet{sch10}, while the distance is measured from the known radius of its
Roche lobe-filling red giant, its well-studied orbit, and its angular radius from optical (K-
band) interferometry \citet{mik16}. We also plot the most reliable 
(i.e. with very well determined maximum brightnesses and $t_2$) novae in the LMC and  M33;
the rapidly recurring nova M31 -12a in M31 \citep{dar16}; the best observed faint-fast novae in M31 \citep{kas11}; 
and the 21 novae from our HST survey of M87.

Our complete sample of M87 novae not only supports the \citet{kas11} claim
that faint, fast novae exist, but triples the sample of such objects, and adds three of the
fastest examples known. These three novae, with $t_2$ of 2.01, 3.72 and 3.75 days, are 
comparable to V597 Pup \citep{hou16}, and the extraordinary recurrent nova M31-12a
in the Andromeda galaxy, which erupts twice every year \citep{hen15} and fades by 2
magnitudes in just 1.65 days \citep{dar16}. A few novae in M31 and elsewhere have
been seen with similar values of $t_2$, but almost always with M = -9.5 to -10 rather than the
values of -7 to -8 observed in M87 and in T CrB.

It is clear from Figure 1 that novae, long believed to be ``standard candles", display
three magnitudes of dispersion in the magnitudes of their MMRD diagram when high cadence, deep CCD sampling 
is used so as not to exclude faint, fast novae. They cannot be reliably used to
measure extragalactic distances, or the distances of newly-discovered Galactic novae. 
This strengthens the similar conclusion reached by \citet{fer03},
albeit based on a smaller and less densely sampled group of nine novae in M49, and by
\citet{kas11} on the basis of the faint, fast novae they detected in M31.

Why did the roughly 100 Galactic, LMC, SMC and M31 novae of the previous century and noted in section 2, provide MMRDs that
yielded a few good extragalactic distances? The surveys that located these objects all suffered from the same incompleteness. The relatively easy-to-find classical novae populate, in zeroth approximation, the upper left and lower right quadrants of MMRD plots. The upper right quadrant is mostly empty ( very slow, very bright novae are rare), while the hard-to-find objects in the lower left quadrant (faint and fast) were almost all missed. The preferential detection of novae in only the upper left and lower right quadrants suggested a spurious correlation - bright objects are preferentially fast and faint objects are slow. Once the lower left quadrant was filled in (with 40\% of all novae) - as has now happened - the apparent correlation vanished.

\clearpage

\begin{sidewaysfigure}
\figurenum{1}
\epsscale{1.0}
\plotone{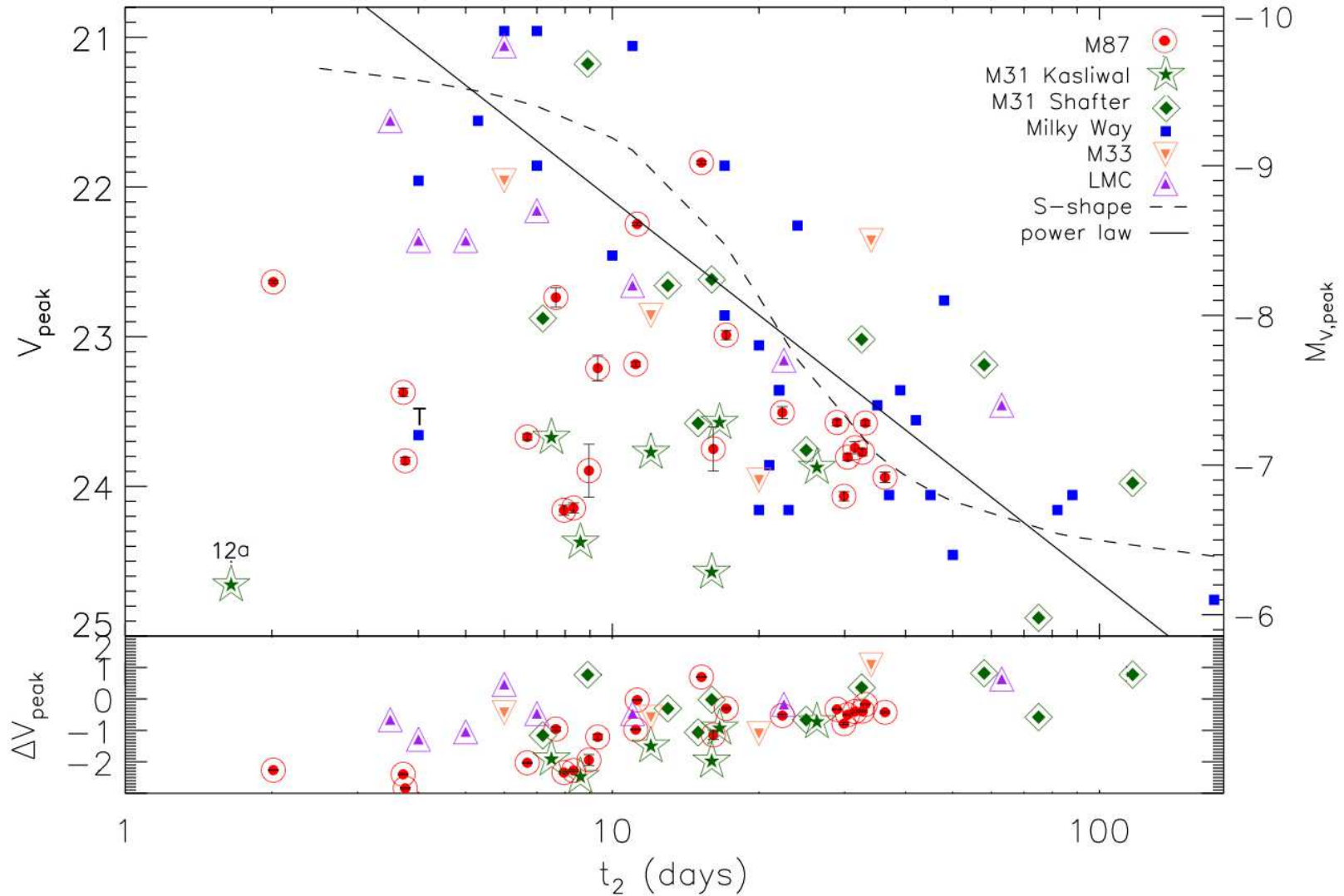}
\caption{Maximum magnitude - Rate of Decline relation (MMRD) for novae with well-defined maxima and $t_2$ in the Milky Way (MW), LMC, M33, M31 and M87. $t_2$ is the time it takes a nova to decline 2 magnitudes from its peak brightness. Filled squares represent MW novae from \citet{dow00}, T denotes the symbiotic nova T CrB \citep{sch10,mik16}, upright triangles are LMC novae from \citet{sha13}, inverted triangles are M33 novae from \citet{sha12}, 12a refers to M31 - 12a \citep{dar16}, open/filled stars denote M31 novae from \citet{kas11} and \citet{sha11}, and open/filled circles denote M87 novae from \citet{sha16}. The M31 data were transformed using transformations from g to V of \citet{jor06}.
The solid and dashed lines represent the best fit power law and S-shaped curves for MW novae \citep{dow00}.  The deviation of each nova from the S-shaped curve in Figure 1 is plotted in the figure's lower panel.}
\end{sidewaysfigure}

\clearpage

\section{Why is there so much scatter in the MMRD plot?}

We have already noted that the mass of the WD in a nova binary is predicted to be an important
parameter in determining how quickly a nova ejects its hydrogen-rich envelope, and thus
how fast it declines from maximum light. This is quantifiable via the 75 self-consistent
models of novae of \citet{yar05}, which not only varied WD mass, but also WD
luminosity and accretion rate onto WDs. \citet{kas11} plotted all these models
in an $M_V$ - timescale diagram, and concluded that ``Some hot and massive white dwarfs
with high accretion rates can result in a faint and fast nova population consistent with the
P60-FasTING sample." We now show that {\it low} accretion rates, and especially {\it low accreted
envelopes masses}, are equally effective at creating faint, fast, {\it non-recurrent} novae on massive WDs.

To clearly separate each of the parameters that determine the location of a nova model
in the $M_V$ - $t_2$ diagram, we superpose onto Figure 1 all 75 of the \citet{yar05} models, color-
coded by WD mass (Figure 2), mass accretion rate (Figure 3), WD core temperature (Figure
4) and total accreted envelope mass (Figure 5). We note that the \citet{yar05} models
calculate $t_3$ (as the timescale of mass-loss $t_{ml}$) rather than $t_2$; we assume that $t_2$ is simply two-thirds of $t_3$. Like 
\citet{kas11}, we assume that novae at maximum luminosity display spectral types close to A5V
to convert the \citet{yar05} maximum model luminosities to $M_V$ .
The models depicted in Figure 2 (and those in \citet{hil16}) predict that any
nova displaying $t_2$ \textless 10 days must contain a WD with a mass in excess of 1.25 $M_{\odot}$. 
Two of \citet{kas11}'s six best-observed M31 novae display $t_2$ \textless 10 days, while nine of
the 21 novae we detected in M87 with well-determined values of $t_2$ do the same. While the
\citet{sha16} survey of M87, spanning 72 days, is ineffective at identifying novae with
$t_2$ longer than 50-60 days, it is clear from their Figure 12 that over 40\% of fast novae display $t_2$ \textless 10 days. 
Such objects are certainly not rare, and reaffirm the claims that WD masses in nova
binaries are much larger, on average, than those in the field \citep{rit91,pag14}.

\begin{sidewaysfigure}
\figurenum{2}
\epsscale{1.0}
\plotone{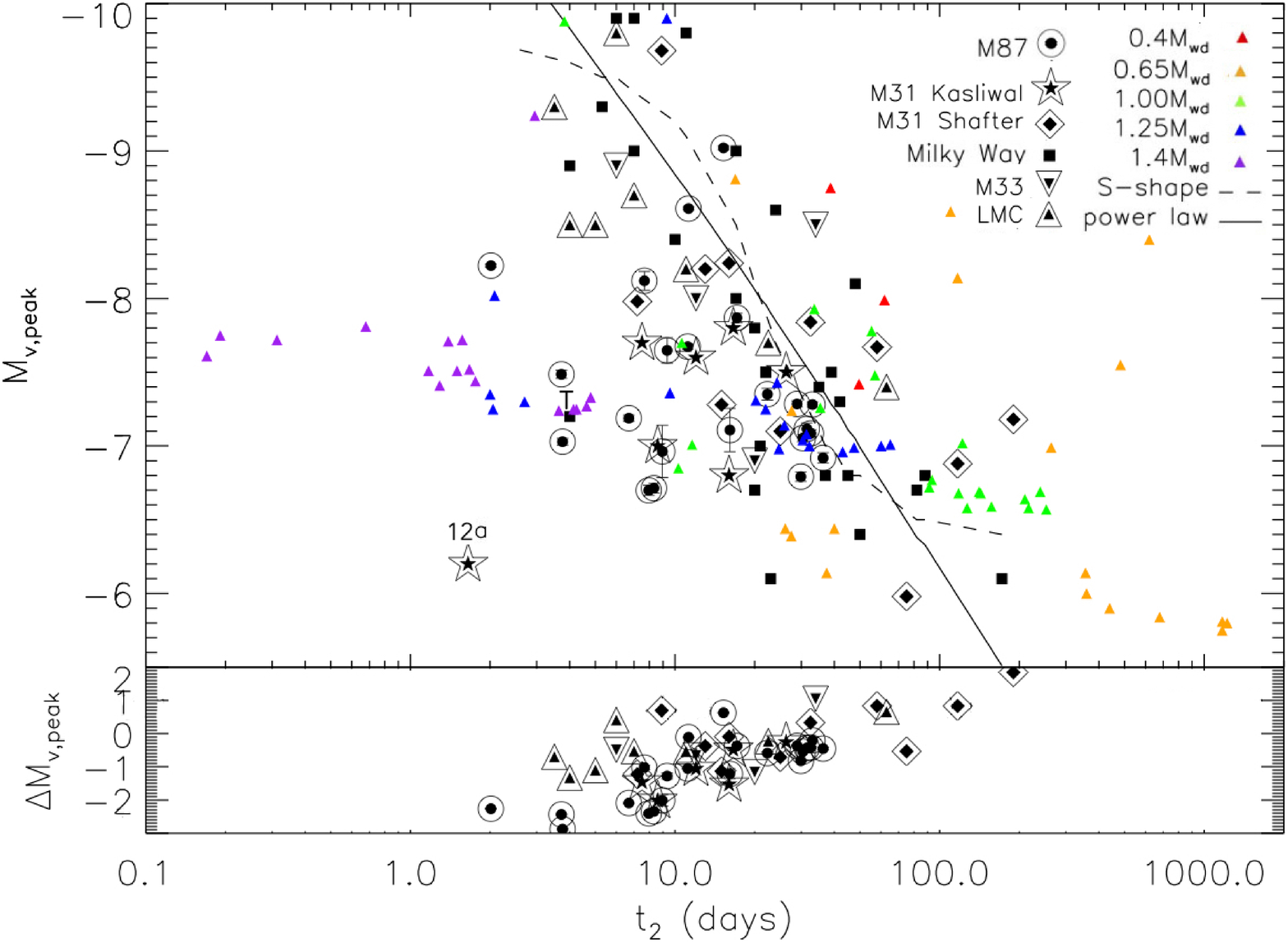}
\caption{Same as Figure 1, but including 75 nova models from \citet{yar05}, as triangles colored according to
the mass of the white dwarf in the nova binary. Larger mass white dwarfs, as explained in the text, correlate with
shorter $t_2$ . The masses of the WDs of the six faint, fast novae discovered by \citet{kas11} (stars in the figure)
are probably in the range 1.0 - 1.25 $M_{\odot}$, while the three fastest novae detected in M87 by \citet{sha16} must
contain white dwarfs close to the Chandrasekhar mass.}
\end{sidewaysfigure}

\clearpage

It is certainly true that varying the rate of mass accretion onto a WD of given mass in
a nova binary can lead to very different outcomes \citep{pac78,pri82}. 
In particular, one might guess that, after WD mass, mass accretion rate is the most
important parameter determining the properties of a nova. In Figure 3 we again replot the
75 nova models of \citet{yar05} on the observational MMRD diagram of Figure 1, but
this time the models are color-coded according to mass accretion rate. In sharp contrast
with Figure 2, where it is apparent that WD mass and $t_2$ are strongly correlated, Figure 3
demonstrates that mass accretion rate and $t_2$ are not correlated at all. Accretion
rates of $10^{-12.3} M_{\odot}$/yr can produce novae with $t_2$ as small as 0.2 days or as large as 500 days.
Peak luminosities, for this same accretion rate, range from $M_V$ = -6.5 to -9.8. Similar large
ranges are seen in both $M_V$ and $t_2$ for other vales of the mass accretion rate.
A similar result is seen in Figure 4, where we replot the 75 Yaron et al. (2005) models
again, but color-coded for WD core temperature (and thus WD luminosity). WD luminosity
by itself plays very little role, if any, in determining the luminosities or decline times of novae.

\begin{sidewaysfigure}
\figurenum{3}
\epsscale{1.0}
\plotone{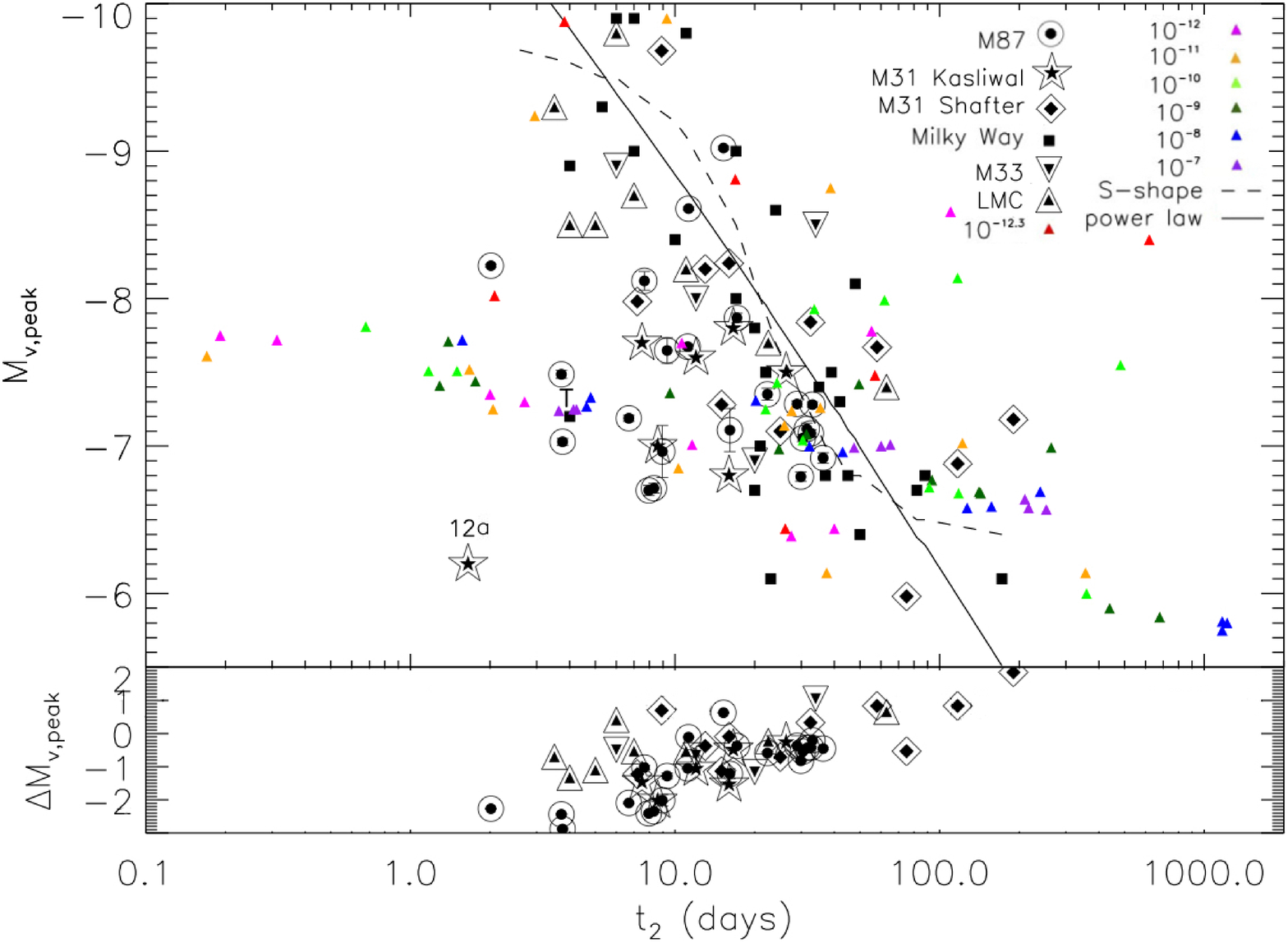}
\caption{Same as Figure 1, but including 75 nova models from \citet{yar05}, as triangles colored according to
the mass accretion rate (assumed constant) onto the white dwarf. Mass accretion rates differing by several orders of
magnitude can produce identical values of $t_2$ or peak luminosity.}
\end{sidewaysfigure}

\clearpage

\begin{sidewaysfigure}
\figurenum{4}
\epsscale{1.0}
\plotone{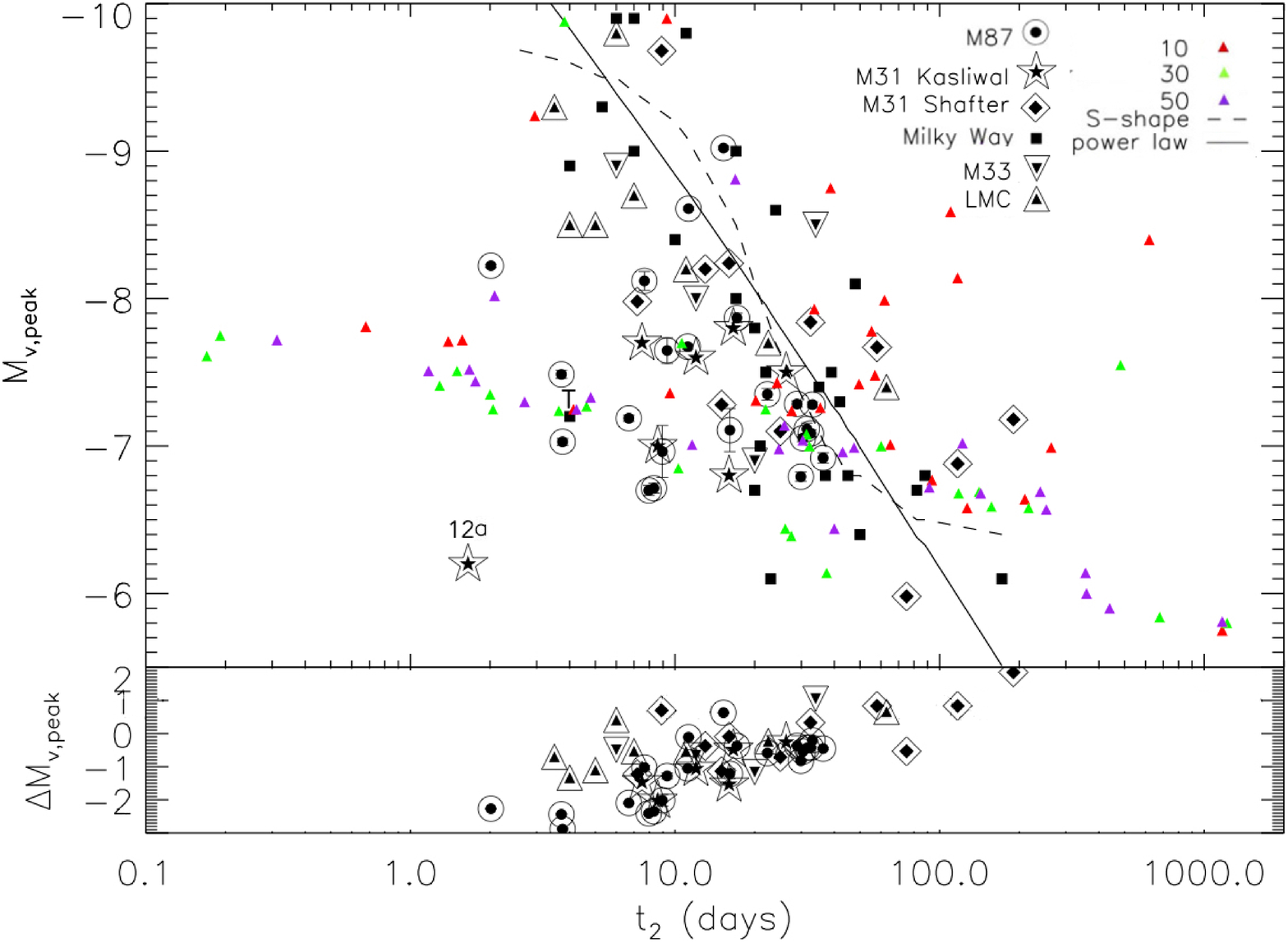}
\caption{Same as Figure 1, but including 75 nova models from \citet{yar05}, as triangles colored according to
the WD core temperature (in units of millions of Kelvins), and hence the luminosity of the white dwarf in the nova binary.}
\end{sidewaysfigure}
\clearpage

The inconclusive results of Figures 3 and 4 are resolved in Figure 5, where we again
plot the \citet{yar05} nova models, but now color-coded according to the mass of the
hydrogen-rich envelope accreted before a nova TNR begins. The correlation between $t_2$ and
accreted envelope mass is evidently much stronger than the correlations of WD luminosity
or mass accretion rate with $t_2$. This is even more obvious in Figure 6, where we plot the
accretion rate, WD temperature, and accreted envelope mass versus $t_2$. 
A useful empirical equation relating these latter two quantities is the least square fit straight line

log $M_{env}$ = 0.825 log ($t_2$) - 6.108 	.										

The underlying reason for the behaviour in Figure 6 is simple: the smallest envelope masses can be ejected
the most quickly, leading to the smallest observed $t_2$.

\begin{sidewaysfigure}
\figurenum{5}
\epsscale{1.0}
\plotone{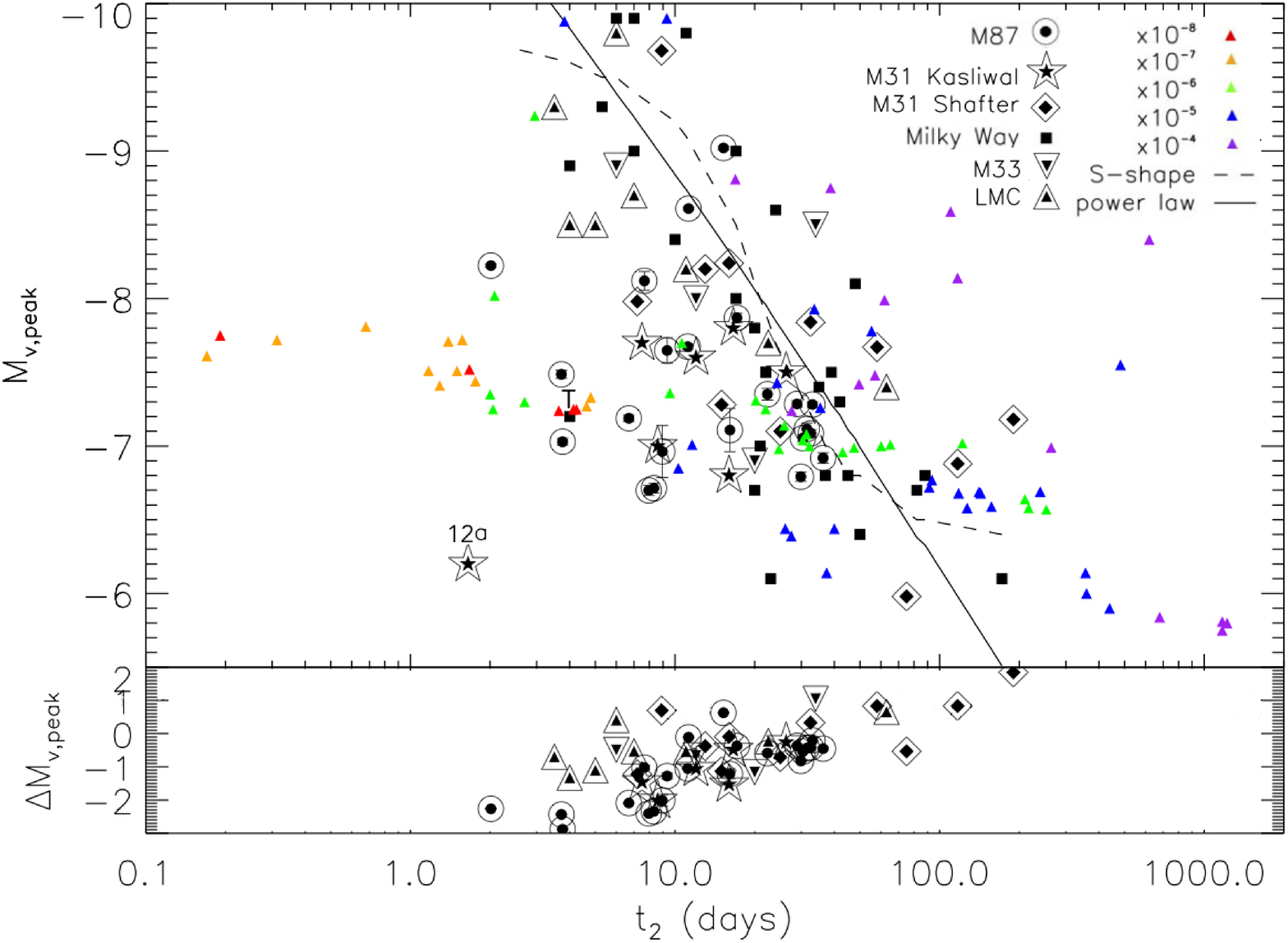}
\caption{Same as Figure 1, but including 75 nova models from \citet{yar05}, as triangles colored according to
the total hydrogen-rich mass accreted onto the white dwarf in the nova binary. The fastest novae
(with smallest $t_2$) have accreted the lowest mass hydrogen-rich envelopes - $10^{-7} - 10^{-8} M_{\odot}$ -
while the slowest novae (largest $t_2$ ) have accreted envelopes 1,000 -10,000 times more massive. From this 
figure, and Figure 2, it is clear that the total accreted envelope mass is as critical a parameter as the WD mass 
in determining the peak luminosity and $t_2$ of a nova.}
\end{sidewaysfigure}

\clearpage

\begin{sidewaysfigure}
\figurenum{6}
\epsscale{1.0}
\plotone{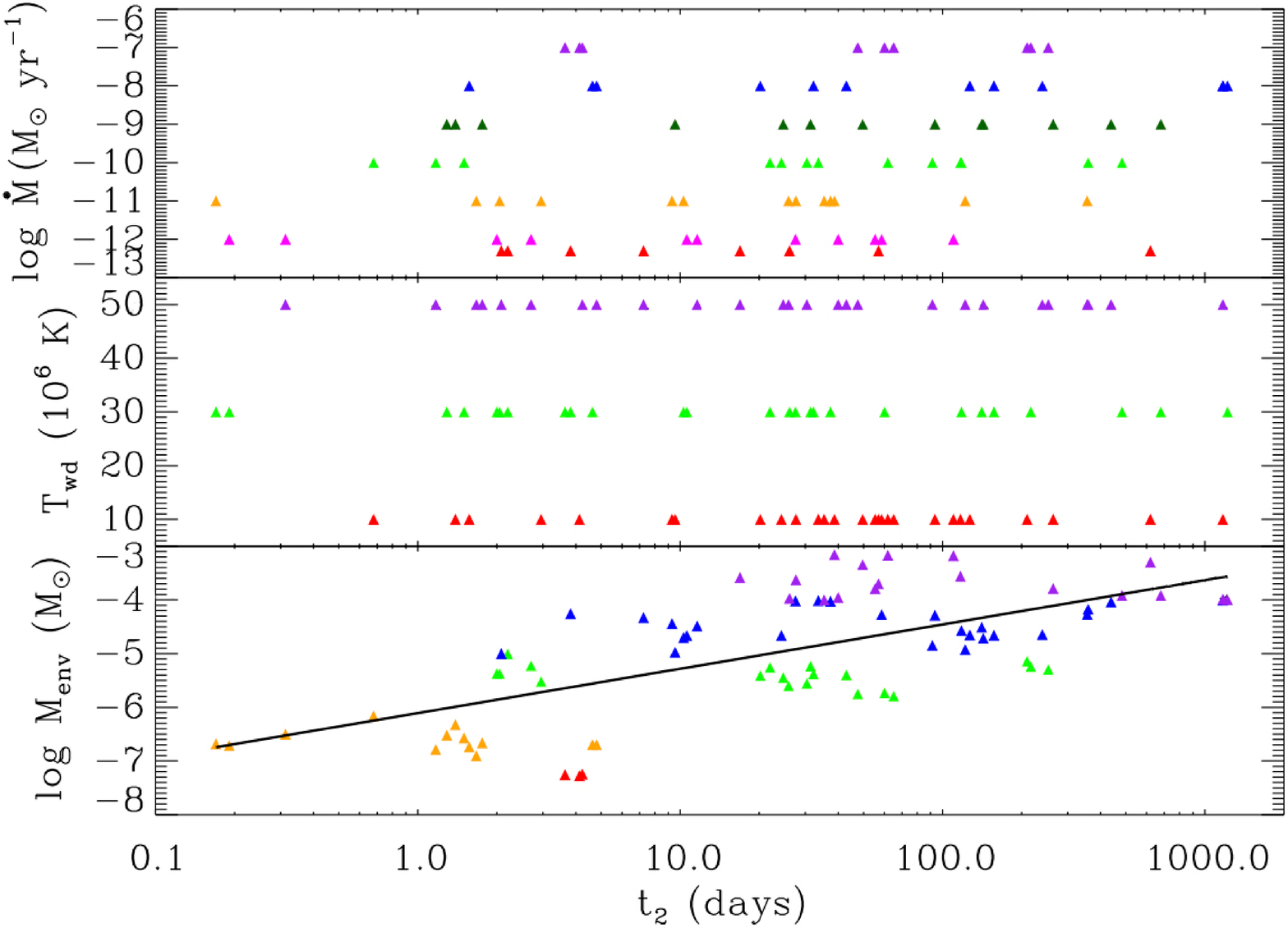}
\caption{Mass accretion rate, WD temperature and accreted envelope mass at the time of eruption, versus $t_2$, for 75
nova models from \citet{yar05}. The colors of triangles in the top, middle and lower panels correspond to the
symbols' keys in Figures 3, 4 and 5. The least-squares fit straight line in the lower panel is discussed in the text.}
\end{sidewaysfigure}
 
\clearpage

\section{A Prediction: the Existence and Detection of Ultrafast Novae}

A strong and testable prediction, overlooked until now, emerges from the models of
\citet{yar05} that is evident in, and physically understandable from Figures 2, 4,
5 and 6. {\it We predict that WDs with masses close to 1.4 $M_{\odot}$, which have 
slowly ($10^{-10} - 10^{-12.3} M_{\odot}$/yr) accreted envelopes of low mass 
($10^{-7} - 10^{-8} M_{\odot}$) can produce novae with $t_2$ as short as 5 hours.}
Simulated light curves of such novae are shown in Figure 4 of \citet{hil14}.

We (rather arbitrarily) define an ultrafast nova as one which displays $t_2$ \textless 1 day. No
such nova has ever been observed, but we maintain that this is entirely due to sampling
bias. Every ground-based survey for extragalactic novae reported in the past century has
employed cadences of a day or longer. Even our own M87 nova survey, which successfully
sampled that galaxy daily for 10 weeks without gaps, was unlikely to  find any nova that
appeared and then faded by 2 magnitudes in less than one day, let alone in 5 hours.

{\it High} accretion rates can accumulate critical envelope masses on timescales as short as
years \citet{yar05}. This is the source of the Recurrent Novae (RNe), which have
massive WDs and inter-eruption intervals of a century or less. Examples include M31-12a
and T CrB. RNe have recently been estimated to comprise 25\% of all novae \citep{pag14}.
We emphasize that (still hypothetical) ultrafast novae are not RNe. Their
low accretion rates must inevitably lead to millennia or longer between their eruptions. If
ultrafast novae are eventually detected, we predict that their ejecta will be very significantly
enhanced in nitrogen relative to the solar value. This is because the long timescale needed
to bring these slowly accreted envelopes to the critical mass for initiation of a TNR will
allow for significant diffusion of hydrogen into the underlying WD \citep{kov85}. This mixing 
enriches the burning envelope of novae by an order of magnitude or more in CNO isotopes, 
which are mostly converted to nitrogen by the TNR.

RNe do not fit the classical nova MMRD \citep{sch10}. But if astronomers are to use MMRD, and to have any confidence in the use of the MMRD for novae discovered in the future, one must be able to distinguish a newly-discovered nova as being a RN or a Classical Nova. (By RN we adopt the conventional definition: an RN erupts at least once per century). \citet{pag14} have exhaustively researched this topic, and demonstrated that the only certain diagnostic of a nova being a RN is observing a second outburst. Thus any newly discovered, fast Galactic or extragalactic nova could be faint and fast (and relatively close), or luminous and fast (and relatively distant). MMRD alone cannot yield a reliable distance for any nova with 
$t_2$ \textless 30 days. Slower novae all display M = $-6.5 \pm 0.5$ mag, but this is 
almost independent of $t_2$ and the MMRD.

Are ultrafast novae rare? A reliable theoretical prediction of the frequency of ultrafast
novae relative to all other novae in a galaxy would involve a population synthesis model
which produces novae from an initial and evolving binary population, and self-consistently
calculates the time-dependent mass transfer rate to the WD in each nova system over that
system's lifetime. This is a challenging problem, far beyond the scope of this paper. A much
simpler approach is to observationally detect ultrafast novae, and measure their relative
frequency amongst all novae.

How might ultrafast novae be detected? The answer is straightforward: via surveys of
nearby galaxies with cadences of order one hour, rather than days. Figure 2 demonstrates
that ultrafast novae should achieve $M_V$ of -6.5 to -7.5, corresponding to 17-18th magnitude in M31.
Detecting such rapid transients, and following them down to 20th magnitude, is within the reach of modern 
CCD cameras attached to 0.5 meter aperture telescopes. The Zwicky Transient Facility (ZTF) will 
utilize a large format camera and the Samuel Oschin 48-inch Palomar Schmidt telescope to begin imaging 
about 3750 square degrees an hour to a depth of 20.5-21 magnitude in 2017. With a 1-hour cadence it should easily discover
ultrafast novae in Local Group galaxies. 
Confirmation, via spectroscopy or narrowband-broadband imaging, can be done in the days
following the detection of rapid transient candidates, as novae remain bright in  H$\alpha\ $
for much longer than they do in continuum light \citep{cia83,nei04}.

\section{Summary and Conclusions}

Nine of 21 well-observed novae in M87 display $t_2$ brightness decline times under 10 days,
and three more have $t_2$ \textless 4 days. These novae are up to 3 magnitudes fainter than predicted by
the MMRD relation, and are similar to the ``faint, fast novae" first detected by \citet{kas11} in M31. 
The fact that these novae are both common and ubiquitous demonstrates
that complete samples of extragalactic novae are not reliable standard candles, and that the MMRD 
should not be used in the era of precision cosmology either for cosmic distance determinations or the distances of Galactic novae.

The \citet{yar05} models of novae explain faint, fast novae as those which occur
on very massive WDs, with very low mass envelopes. Low mass envelopes that were accreted
quickly lead to RNe. We predict that those accreted slowly yield (previously overlooked) ultrafast novae
which brighten and fade by 2 magnitudes in under 1 day. Such ultrafast novae are also predicted to display large
nitrogen enhancements relative to the solar value. We predict that surveys of M31 and other
nearby galaxies with cadences of order 1 hour will reveal these novae, even with modest-sized
telescopes.

We gratefully acknowledge the support of the STScI team responsible for ensuring timely
and accurate implementation of our M87 program. Support for program \#10543 was 
provided by NASA through a grant from the Space Telescope Science Institute, which is operated
by the Association of Universities for Research in Astronomy, Inc., under NASA contract
NAS 5-26555. This research has been partly supported by the Polish NCN grant DEC-
2013/10/M/ST9/00086. MMS gratefully acknowledges the support of Ethel Lipsitz and the
late Hilary Lipsitz, longtime friends of the AMNH Astrophysics department.

\end{document}